\documentclass[
  journal=pasa,
  manuscript=research-paper, 
  year=2020,
  volume=37,
]{cup-journal}

\usepackage{microtype,siunitx,booktabs}
\title{The thermal history of the intergalactic medium at \boldmath $3.9 \leq z \leq 4.3$}

\author{T. Ondro}
\affiliation{Department of Technology and Automobile Transport, Faculty of AgriSciences, Mendel University in Brno, Zem\v{e}d\v{e}lsk\'{a} 1, 613 00 Brno, Czech Republic}
\alsoaffiliation{Institute of Physics, Faculty of Science, Pavol Jozef \v{S}af\'{a}rik University in Ko\v{s}ice, Park Angelinum 9,
040 01 Ko\v{s}ice, Slovakia}
\email[T. Ondro]{tomas.ondro@mendelu.cz}

\author{R. G\'{a}lis}
\affiliation{Institute of Physics, Faculty of Science, Pavol Jozef \v{S}af\'{a}rik University in Ko\v{s}ice, Park Angelinum 9,
040 01 Ko\v{s}ice, Slovakia}

\doi{10.1017/pasa.2020.32}

\received {dd Mmm YYYY}
\revised  {dd Mmm YYYY}
\accepted {dd Mmm YYYY}
\published{22 September 2020}


\begin{document}

\begin{abstract}
A new determination of the temperature of the intergalactic medium over $3.9 \leq z \leq 4.3$ is presented. We applied the curvature method on a sample of 10 high resolution quasar spectra from the Ultraviolet and Visual Echelle Spectrograph on the VLT/ESO. We measured the temperature at mean density by determining the temperature at the characteristic overdensity, which is tight function of the absolute curvature irrespective of $\gamma$. Under the assumption of fiducial value of $\gamma = 1.4$, we determined the values of temperatures at mean density $T_{0} = 7893^{+1417}_{-1226}$\,K and $T_{0} = 8153^{+1224}_{-993}$\,K for redshift range of $3.9 \leq z \leq 4.1$ and $4.1 \leq z \leq 4.3$, respectively. Even though the results show no strong temperature evolution over the studied redshift range, our measurements are consistent with an intergalactic medium thermal history that includes a contribution from He\,{\sc ii} reionization.
\end{abstract}

\section{INTRODUCTION}
\label{sec:intro}
The thermal state of the gas in the intergalactic medium (IGM) is an important characteristic describing the baryonic matter in the Universe \citep{Lidz2010}. \citet{HuiGnedin} showed that the temperature-density relation of the photoionized IGM in the low-density region can be well-approximated by the following equation
\begin{equation}
T=T_{0}\Delta^{\gamma-1},
\label{asymptot_td_relation2}
\end{equation}
where $T_{0}$ is the temperature at the mean density, $\Delta$ is the overdensity and $(\gamma - 1)$ is a power-law index. 

The standard model assumes that the evolution of the IGM has passed through two major reheating events. At first, the reionization of hydrogen (H\,{\sc i} $\rightarrow$ H\,{\sc ii}) occurred, and it is normally assumed that helium is singly ionized (He\,{\sc i} $\rightarrow$ He\,{\sc ii}) along with H\,{\sc i}. This process should be completed at the redshift $z\sim6$ \citep{Bouwens2015}. Then, the IGM cooled and is reheated again during the He\,{\sc ii} reionization phase. This process is expected to be completed at the redshift $z\sim2.7$ and can be characterized by three phases \citep{Worseck2011}:

\begin{enumerate}
\item He\,{\sc iii} "bubble" growth around quasars (QSOs) with redshifts $z_{\text{em}} \geq 4$,
\item overlap of the He\,{\sc iii} zones around more abundant QSOs at $z_{\text{reion}} \sim 3$,
\item gradual reionization of remaining dense He\,{\sc ii} regions.
\end{enumerate} 

In recent years, an attention has been paid to characterizing the $T-\rho$ relation of the IGM around $z \sim 3$ \citep{Schaye2000, Rorai2018, Hiss_2018, Telikova_2018, Telikova_2019, Walther2019, gaikwad2021}. 

However, in case of the higher redshifts ($z > 4$), absorption features start to become strongly blended \citep{Becker2011}. Due to this, most studies are based on the Ly-$\alpha$ flux power spectrum \citep{Garzilli2017, Irsic2017, Walther2019, Boera_2019}. \citet{Becker2011} used the curvature statistic, which does not require the decomposition of the Ly-$\alpha$ forest into individual spectral lines. There is only one study treating the Ly-$\alpha$ forest as a superposition of discrete absorption profiles \citep{Schaye2000} at $z \sim 4$.

\begin{table*}
\begin{threeparttable}
	\centering
	\caption{List of QSOs whose spectra were used in this study. The S/N ratio was calculated according to \citet{Stoehr2008} for the spectral regions where the absorbers were parameterized.}
	\label{tab:QSO_data}
\begin{tabular}{l c c c c c}\toprule
		Object & R.A. (J2000) & Dec. (J2000) & $z_{\rm{em}}$ & S/N & ESO Program IDs\\ \midrule
J020944+051713	&	02:09:44.61	&	+05:17:13.6	&	4.184	&	14	&	69.A-0613(A)	\\
J024756-055559	&	02:47:56.56	&	-05:55:59.1	&	4.238	&	12	&	71.B-0106(B)	\\
J030722-494548	&	03:07:22.90	&	-49:45:48.0	&	4.728	&	18	&	60.A-9022(A)	\\
J095355-050418	&	09:53:55.74	&	-05:04:18.9	&	4.369	&	13	&	072.A-0558(A)	\\
J120523-074232	&	12:05:23.11	&	-07:42:32.7	&	4.695	&	26	&	166.A-0106(A),66.A-0594(A),71.B-0106(A)	\\
J144331+272436	&	14:43:31.16	&	+27:24:36.7	&	4.43	&	14	&	072.A-0346(B),077.A-0148(A),090.A-0304(A)	\\
J145147-151220	&	14:51:47.03	&	-15:12:20.2	&	4.763	&	28	&	166.A-0106(A)	\\
J201717-401924	&	20:17:17.12	&	-40:19:24.1	&	4.131	&	11	&	71.A-0114(A)	\\
J215502+135825	&	21:55:02.01	&	+13:58:25.8	&	4.256	&	11	&	65.O-0296(A)	\\
J234403+034226	&	23:44:03.11	&	+03:42:26.7	&	4.239	&	10	&	65.O-0296(A)	\\

\bottomrule
\end{tabular}
\end{threeparttable}
\end{table*}

The aim of this work is to study the thermal history of the IGM at $3.9 \leq z \leq 4.3$ using curvature statistics. Besides, we compare our measurements with the $T_{0}$ evolution predicted by widely used spatially homogeneous UVB models of \citet{Haardt_2012}, \citet{Onorbe2017}, \citet{Khaire}, \citet{Puchwein2019}, and \citet{Faucher2020}. To be more specific, we compare results with the rescaled models, same as in the study by \citet{gaikwad2021}. The results in the aforementioned study are consistent with the relative late He\,{\sc ii} reionization in the models of \citet{Onorbe2017} and \citet{Faucher2020}, in which the mid-point of the He\,{\sc ii} reionization is at $z_{\rm mid} \sim 3$. On the other hand, in case of the other compared models \citep{Haardt_2012, Khaire, Puchwein2019}, we can observe stronger temperature evolution in the redshift range of $3.8 < z < 4.4$. Additional impetus could be that even there is good agreement between theory and observations for the temperature evolution of the IGM at $\sim 3$, there is still a lack of data at higher redshifts.

The article is organized as follows: Section 2 contains the basic information about the observational data used in this work. The curvature method and the summary of the analysis together with the sources of uncertainties are described in Section 3. A description of the used simulations and an explanation of the generation of the simulated spectra are given in Section 4. In Section 5, we present our results and their comparison with the previously published ones and with $T_{0}$ evolution predicted by widely used spatially homogeneous UVB models. Our conclusions are given in Section 6.

\section{OBSERVATIONS}
\label{sec:Observ}
\begin{figure*} 
\centering
	\includegraphics[width=0.85\columnwidth]{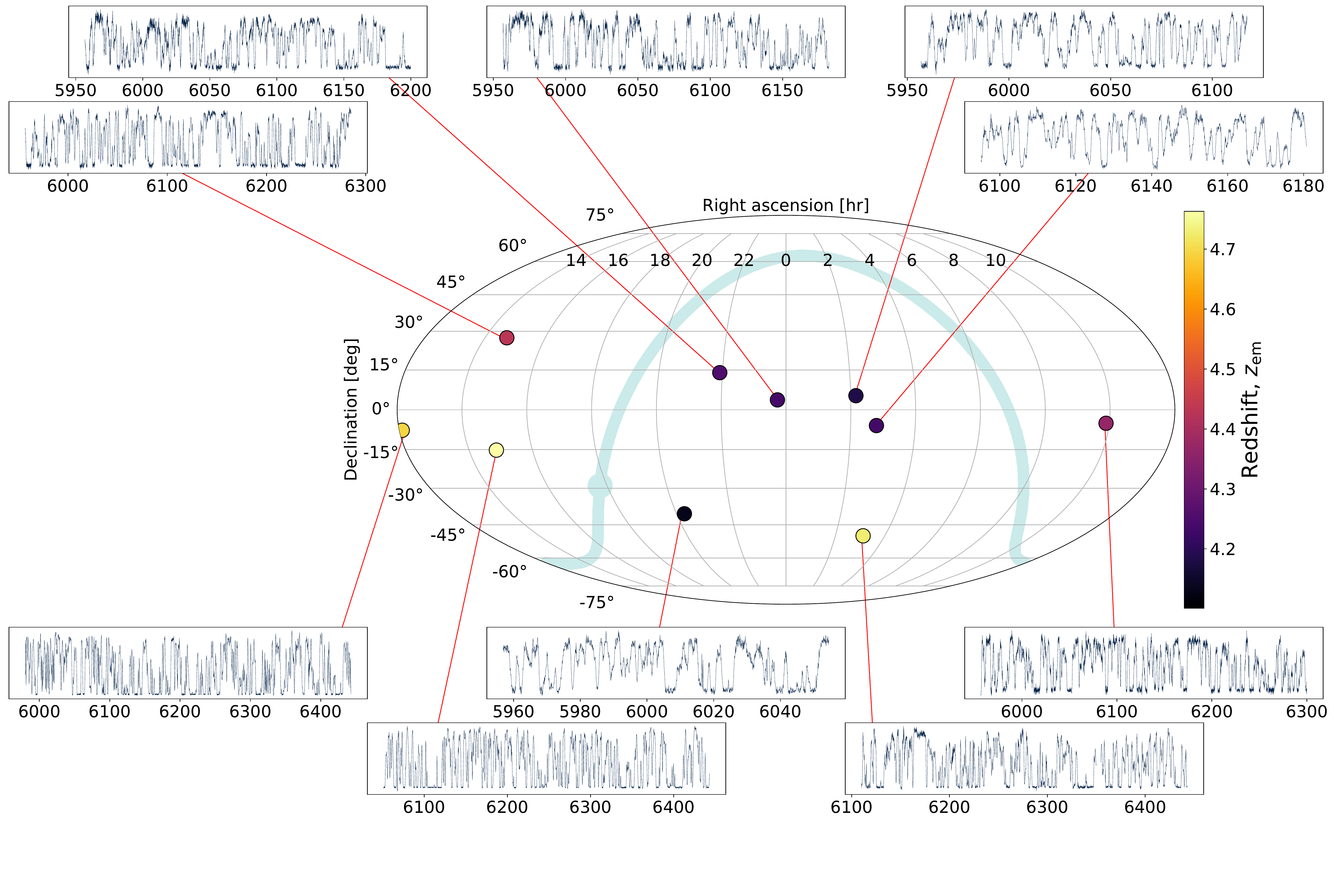}
    \caption{The coverage of the dataset in which each spectrum represents the Ly-$\alpha$ redshift range for individual QSOs from our sample.}
    \label{fig:qso_redshift}
\end{figure*}

In this study, we used a sample of QSO spectra (Tab. \ref{tab:QSO_data}) obtained by the Ultraviolet and Visual Echelle Spectrograph (UVES) on the VLT/ESO \citep{UVES_dataset}. The UVES Spectral Quasar Absorption Database contain fully reduced, continuum-fitted high-resolution spectra of quasars in the redshift range $0 < z < 5$. The spectral data has nominal resolving power $R_{\rm nom}\simeq 50 000$ and dispersion of $2.5\,\rm km\,s^{-1}\,\rm pixel^{-1}$. From the whole dataset we selected only spectra that meet the following criteria:
\begin{enumerate}
    \item the sightline partially or fully contains the Ly-$\alpha$ forest in the redshift range of $3.9 \leq z \leq 4.3$. 
    To be more specific, we focused on the spectral region of rest-frame wavelengths 1050\,\AA\ -- 1180\,\AA\, inside the Ly-$\alpha$ forest. This is the same range used in \citet{Delabrouille2013,Hiss_2018,Walther2018,Ondro2021} and is considered a conservative choice for the Ly-$\alpha$ forest region.
   \item the signal-to-noise (S/N) ratio of the spectrum is higher than 10 in the studied spectral region.
\end{enumerate}
We used a sample of 10 QSO spectra, which fulfills above criteria, where the coverage of the analysed QSO spectra is shown in Fig. \ref{fig:qso_redshift}.

Note that the Ly-$\alpha$ absorbers for which $\log{N_{\text{H\,{\sc i}}}} \geq 20$ (damped Ly-$\alpha$ systems) were identified by eye and excluded from the analysis. In this case, the excluded part of the spectrum was chosen to enclose the region between the points where the damping wings reached a value below 0.9 within the flux error. This value was chosen because the flux only occasionally reaches the continuum value. The spectral intervals with bad pixels were masked and cubically interpolated.  

\section{CURVATURE METHOD}
\label{sec:curvature}
In this work, we applied the curvature method to obtain new, robust determinations of the IGM temperature at redshift range of $3.9 \leq z \leq 4.3$. The curvature $\kappa$ is defined as \citep{Becker2011}
\begin{equation}
\kappa \equiv \frac{F''}{[1 + (F')^{2}]^{3/2}},
\end{equation}
where the $F' = {\rm d}F / {\rm d}v$ and $F'' = {\rm d}^{2}F / {\rm d}v^{2}$ are the first and second derivatives of the flux field with respect to velocity, respectively. The greatest advantage of this method is that it does not require the decomposition of the Ly-$\alpha$ forest into individual lines. This is useful mainly in the higher redshifts, where absorption features start to become strongly blended. 

Due to the reproducibility, we describe the basic steps of the curvature calculation in  \ref{Sec:app1}.

\subsection{Sources of Uncertainties}
As already shown, $\kappa$ is easy to compute and can be evaluated on a pixel-by-pixel basis \citep{Becker2011}. Before using it, however, several issues need to be addressed, which are described below.

\subsubsection{Noise}
The curvature can be affected by the finite S/N of the spectra. To solve this difficulty \citet{Becker2011} and \citet{Boera2014} fitted the $b$-spline to the flux, and then compute the curvature from the fit. In this study we used the same approach as \citet{gaikwad2021}, and we smoothed the flux using the Gaussian filter of FWHM $\sim 10$ km s$^{-1}$. The similar approach was used also in \citet{Padmanabhan2015}. 

\subsubsection{Continuum}
In general, for spectra with high-resolution and high $S/N$, the continuum is fitted by locally connecting apparent absorption-free spectral regions. However, this approach depends on the average line density, and thus on the redshift. At higher redshifts (typically $z > 4$), severe blendings makes it hard or even impossible to identify the unabsorbed spectral regions. Therefore, a polynomial with typically 3 to 5 degrees of freedom for the region from Ly-$\alpha$ to Ly-$\beta$ can be used. This approach can produce the statistical uncertainty of the continuum placement exceeding 7\% \citep{Becker2007,UVES_dataset}. To circumvent the continuum issue, we re-normalized both, the real data and also the simulations. For each 20\,Mpc/$h$ section, we divided the flux by the maximum value of smoothed flux in that interval. 

\subsubsection{Metal Lines}
\label{metal_rejection}
\begin{figure}
\centering
	\includegraphics[width=\columnwidth]{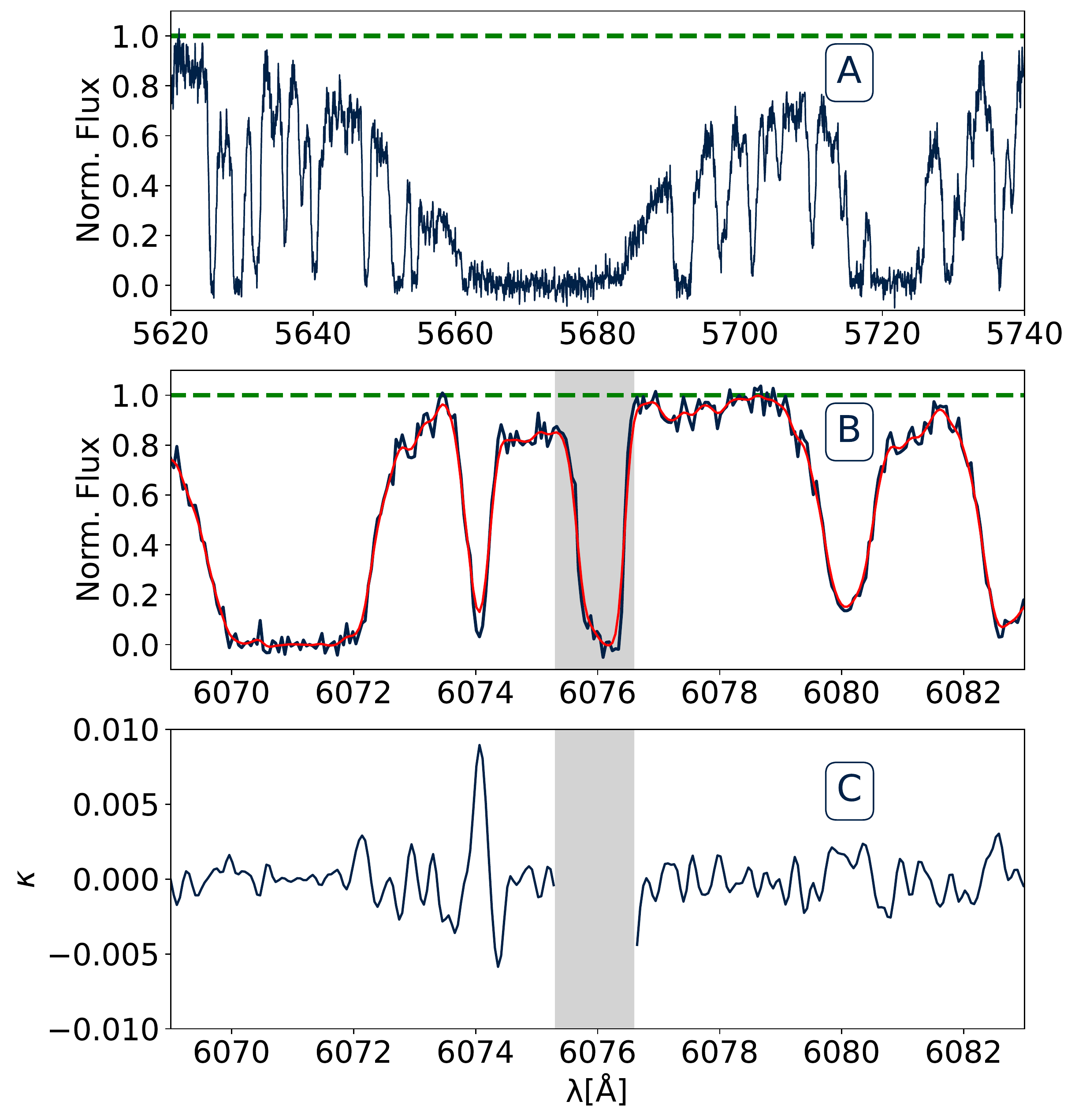}
    \caption{An example of the adopted procedure for rejecting metal lines based on the DLA system at $z \approx 3.666$ (A) in the spectrum of the quasar QSO J020944 + 051713. The green dashed line and red solid line represents the continuum level and result of the smoothing the flux using the Gaussian filter, respectively. The shaded region (B) demonstrates the excluded part of curvature field (C) due to the contamination of the region by metal absorption (O\,{\sc i} 1302) .}
    \label{fig:Metal_lines}
\end{figure}

It is well known that the Ly-$\alpha$ forest is contaminated by metal lines, which are a potentially serious source of systematic errors \citep{Boera2014}. These lines are usually associated with the strong \text{H\,{\sc i}} absorption. For this reason, we visually inspected the studied spectra to identify damped Ly-$\alpha$ (DLA) and sub-DLA systems, for which the redshifts were determined. In this case, the associated metal lines redward of the Ly-$\alpha$ emission peak of the QSO help with the proper determination of the DLA redshift (see Fig. \ref{fig:Metal_lines}). If the redshifts were known, the other metal lines (Tab. \ref{tab:Metal_transitions}) were determined based on their characteristic $\Delta \lambda$.

It is worth noting that in the case of other metal-absorption systems not associated with the DLA, we used the doublet metal lines (typically Si\,{\sc iv}, C\,{\sc iv}) to determine the redshift of metal-absorption systems. 

Following these steps we firstly compute the curvature. Then, based on the determined redshifts, the expected wavelength of the metal lines were calculated. Finally, we excluded a region of the curvature field, which corresponds to the 30 km s$^{-1}$ in each direction around each potential metal line, so that the metal lines did not affect the results of our analysis.

\begin{table}
\begin{center}
	\caption{List of metal lines included in our semi-automatic rejection procedure with their oscillator strength $f$.}
	\label{tab:Metal_transitions}
	\begin{tabular}{lccc} 
		\hline
		\textbf{Absorber} & \textbf{$\lambda_{\text{rest}}$ [\AA]} & {\boldmath{$f$}} & \textbf{Reference} \\
		\hline
O\,{\sc vi}	    &	1031.9261	&	0.13290	&	1	\\
C\,{\sc ii}	    &	1036.3367	&	0.12310	&	1	\\
O\,{\sc vi}	    &	1037.6167	&	0.06609	&	1	\\
N\,{\sc ii}	    &	1083.9900	&	0.10310	&	1	\\
Fe\,{\sc iii}	&	1122.5260	&	0.16200	&	2	\\
Fe\,{\sc ii}	&	1144.9379	&	0.10600	&	3	\\
Si\,{\sc ii}	&	1190.4158	&	0.25020	&	1	\\
Si\,{\sc ii}	&	1193.2897	&	0.49910	&	1	\\
N\,{\sc i}	    &	1200.2233	&	0.08849	&	1	\\
Si\,{\sc iii}	&	1206.5000	&	1.66000	&	1	\\
N\,{\sc v} 	    &	1238.8210	&	0.15700	&	1	\\
N\,{\sc v}	    &	1242.8040	&	0.07823	&	1	\\
Si\,{\sc ii}	&	1260.4221	&	1.00700	&	1	\\
O\,{\sc i}	    &	1302.1685	&	0.04887	&	1	\\
Si\,{\sc ii}	&	1304.3702	&	0.09400	&	4	\\
C\,{\sc ii}	    &	1334.5323	&	0.12780	&	1	\\
C\,{\sc ii}*	&	1335.7077	&	0.11490	&	1	\\
Si\,{\sc iv}	&	1393.7550	&	0.52800	&	1	\\
Si\,{\sc iv}	&	1402.7700	&	0.26200	&	1	\\
Si\,{\sc ii}	&	1526.7066	&	0.12700	&	5	\\
C\,{\sc iv}	    &	1548.1950	&	0.19080	&	1	\\
C\,{\sc iv}	    &	1550.7700	&	0.09522	&	1	\\
Fe\,{\sc ii}	    &	1608.4511	&	0.05800	&   2	\\
Al\,{\sc ii}	&	1670.7874	&	1.88000	&	1	\\
Al\,{\sc iii}	&	1854.7164	&	0.53900	&	1	\\
Al\,{\sc iii}	&	1862.7895	&	0.26800	&	1	\\
Fe\,{\sc ii}	    &	2344.2140	&	0.11400	&	2	\\
Fe\,{\sc ii}	    &	2374.4612	&	0.03130	&	2	\\
Fe\,{\sc ii}	    &	2382.7650	&	0.32000	&	2	\\
Fe\,{\sc ii}	    &	2586.6500	&	0.06910	&	2	\\
Fe\,{\sc ii}	    &	2600.1729	&	0.23900	&	2	\\
Mg\,{\sc ii}	&	2796.3520	&	0.61230	&	6	\\
Mg\,{\sc ii}	&	2803.5310	&	0.30540	&	6	\\
Mg\,{\sc i}	&	2852.9642	&	1.81000	&	1	\\
		\hline
	\end{tabular}\\
	\end{center}
References: (1) \cite{Morton1991}, (2) \cite{Prochaska2001}, (3) \cite{Howk2000}, (4) \cite{Tripp1996}, (5) \cite{Schectman1998}, (6) \cite{Verner1996}.
\end{table}

\subsection{Summary of method}
The whole analysis can be summarized as follows:
\begin{enumerate}
\item We divided the spectra into 20\,Mpc/$h$ sections to directly match the box size of the simulated spectra.
\item The flux field is smoothed using the Gaussian filter of FWHM $\sim 10$ km s$^{-1}$.
\item We re-normalized the flux, which was already normalized by the broader spectral range fit of the continuum, by dividing the flux of each section by the maximum value of the smoothed flux field in that interval.
\item The curvature $\kappa$ is determined and only pixels, in which the value of the re-normalized flux $F^{R}$ falls in the range of $0.1 \leq F^{R} \leq 0.9$ are taken into consideration. The lower value was chosen due to the fact that the saturated pixels do not contain any information on the temperature. Using the higher threshold we exclude the pixels with flux near the continuum.
\item We masked the metal lines.
\item In the case of real QSO spectra, we joint the curvature values from all of the 20 Mpc/h sections and determined the median of the $<|\kappa|>$ from the 5000 moving blocks bootstrap realizations.
\item We also applied the same procedure in case of the simulations, which were prepared for the analysis according to the procedure described in the next section. Note that the metal contamination in case of the simulations is not considered.
\item Finally, the temperature $T(\overline{\Delta})$ is calculated by interpolating the $T(\overline{\Delta}) - \log{<|\kappa|>}$ relation based on the simulations to the value of $\log{<|\kappa|>}$ determined from the data.
\end{enumerate}
It is worth noting that all uncertainties in this study correspond to the 2.5th and 97.5th percentiles of distribution based on the bootstrap realizations.

\section{SIMULATIONS}
\label{sec:sim}
\begin{figure}[t]
\centering
	\includegraphics[width=0.8\columnwidth]{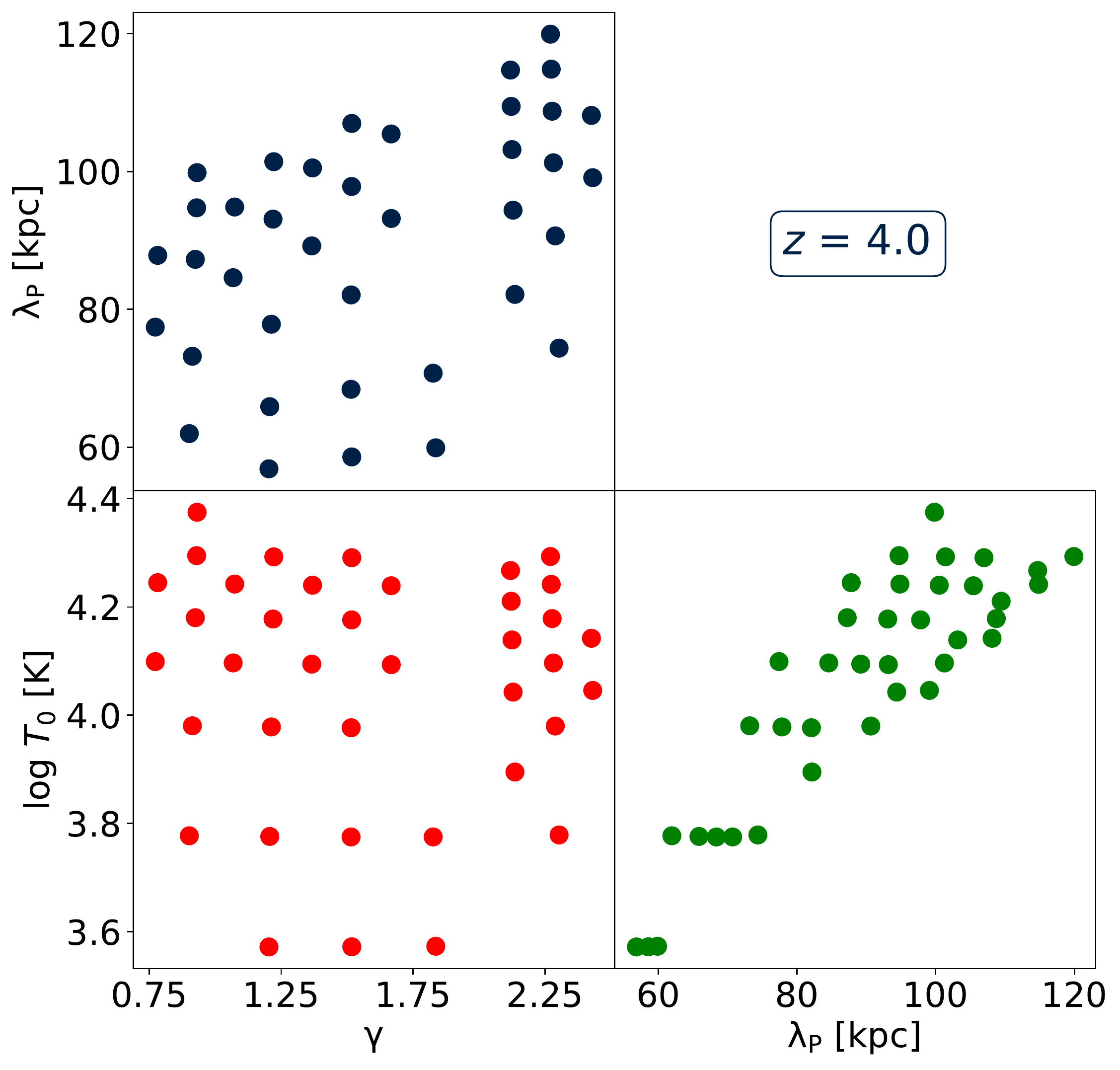}\\
	\vspace{3em}
    \includegraphics[width=0.8\columnwidth]{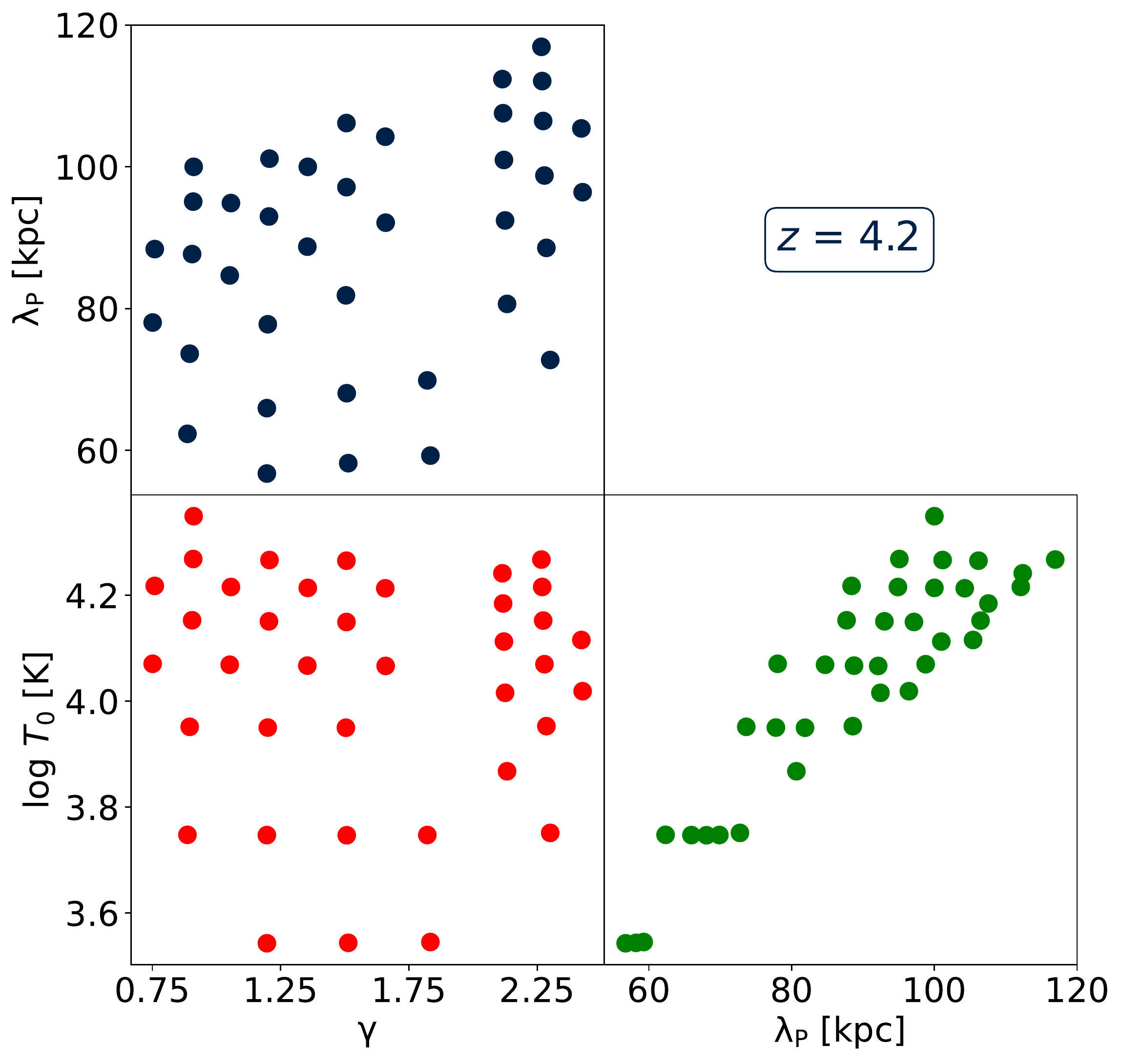}
	\caption{Combinations of $T_{0}$, $\gamma$ and $\lambda_{\rm P}$.}
    \label{fig:sim_param_comb}
\end{figure}

In this study, we used a part of the THERMAL\footnote{thermal.joseonorbe.com} suite, which consists of $\sim$ 70 Nyx hydrodynamical simulations with different thermal histories on a~box size $L_{\rm box} = 20$\,Mpc/$h$ and 1024$^3$ cells \citep[see details in][]{Onorbe2017, Hiss_2018,Hiss_2019}. From the whole dataset, we chose a subset of 38 simulation snapshots at $z = 4.0$ and also at $z = 4.2$, with different combinations of underlying thermal parameters $T_{0}$, $\gamma$ and pressure smoothing scale $\lambda_{P}$, which satisfy a spacing threshold
\begin{equation}
    \sqrt{ \left( \frac{T_{i} - T_{j}}{\text{max}(T) - \text{min}(T)} \right)^2 +  \left( \frac{\gamma_{i} - \gamma_{j}}{\text{max}(\gamma) - \text{min}(\gamma)} \right)^2} \geq 0.1.
\end{equation}
This condition was based on the fact that some of the models have close values of their thermal parameters. This is the similar approach that was used by \citet{Walther2019}. The final subsets of simulations, with different combinations of thermal parameters are depicted in the Fig. \ref{fig:sim_param_comb}.

Note that the parameters $T_0$ and $\gamma$ were determined from the simulations by fitting a power-law temperature-density relation to the distribution of gas cells using linear least squares method as described in \citet{Lukic2015}. In order to determine the $\lambda_{P}$, the approach present in \citep{Kulkarni2015} was used. The cosmological parameters used in the simulations were based on the results of the \textit{Planck} mission \citep{Planck2014}: $\Omega_{\Lambda} = 0.6808$, $\Omega_{\rm m} = 0.3192$, $\sigma_{8} = 0.826$, $\Omega_{\rm b} = 0.04964$, $n_{\rm s} = 0.96$, and $h = 0.6704$. 

\subsection{Skewer generation}
\label{Skewer_Generation}
In the next step, for each model that fulfils the aforementioned conditions, we transformed the Ly-$\alpha$ optical depth $(\tau)$ skewer into the corresponding flux skewer $F$ according to the equation $F = F_{c} \exp{(-A_r \tau)}$, where continuum flux $F_{c}$ was set up to unity and $A_r$ is the scaling factor, which allows us to match the lines of sight to observed mean flux values. Its value can be determined by comparing of the mean flux of the simulations with observational mean flux. In this study, we used the value that corresponds to the mean flux evolution presented in \citet{Onorbe2017}, which is based on accurate measurements of \citet{Fan2006, Becker2007, Kirkman2007, Faucher2008}, and \citet{Becker2013}. It is worth noting, that the mean flux normalization is computed for the full snapshot.

\subsection{Modeling noise and resolution}
To create mock spectra, we added effects of resolution and noise to the simulated skewers. The magnitude of both effects was adjusted so that the mock spectra corresponded as closely as possible to the observed ones. 
Note that in case of the real data, we divided the spectra into 20\,Mpc/$h$ sections to directly match the box size of the simulated spectra, and calculate the S/N of each section. Then, we match the S/N of the simulated spectra with the section of the QSO spectra.

\section{RESULTS AND DISCUSSION}
\begin{figure}
    \centering
    \includegraphics[width=0.75\columnwidth]{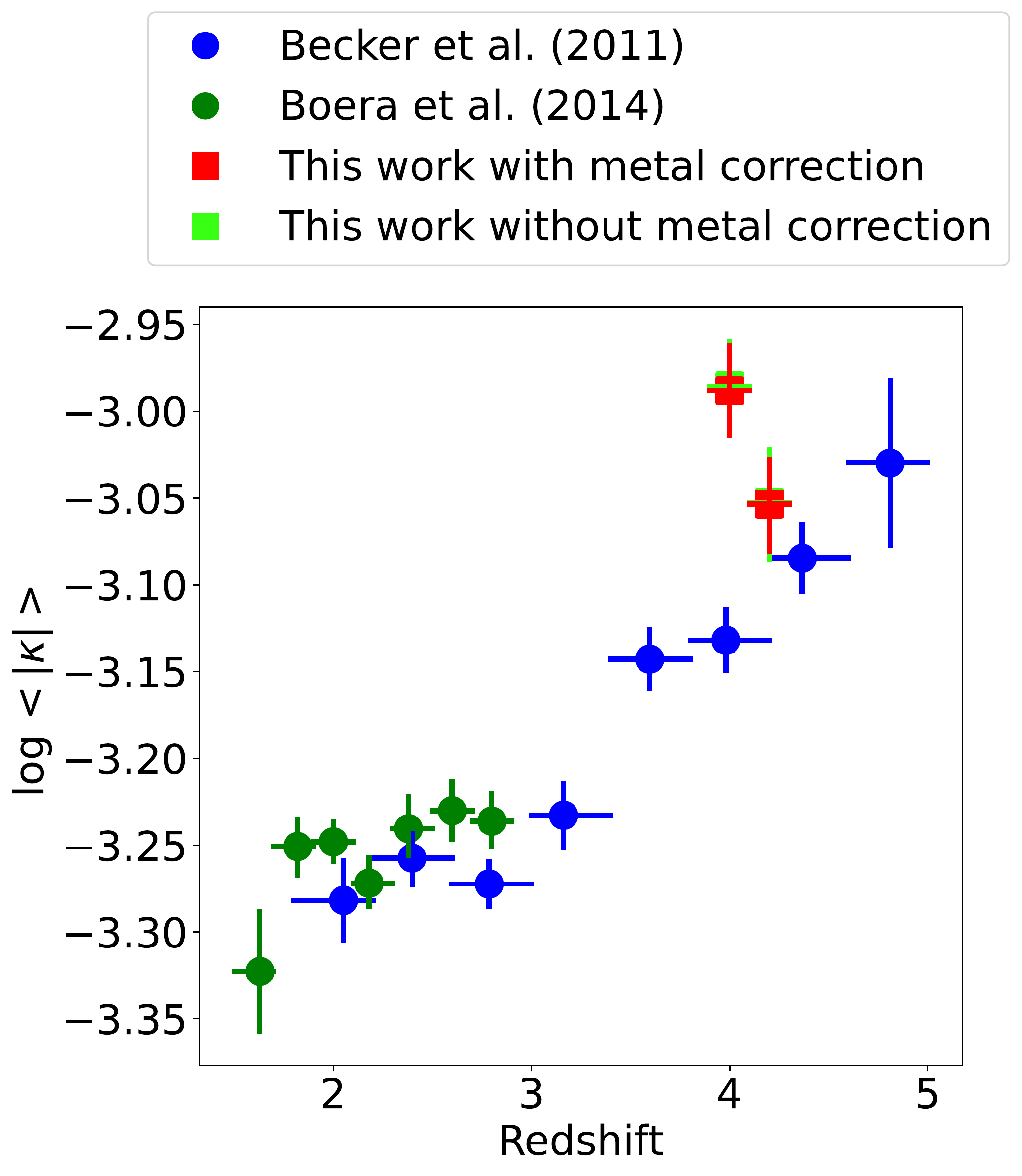}
    \caption{Curvature measurements from the observational QSO spectra.}
    \label{fig:curvature}
\end{figure}

In this section, we present the measured values of the temperature at the mean density by determining the temperature at the characteristic overdensity, which is a tight function of the absolute curvature irrespective of the $\gamma$. The final results for the curvature measurements from the real QSO spectra are shown in Fig. \ref{fig:curvature}. The results show that there is a small difference between the values of curvature with and without metal correction. However, there is a~significant difference of $\log{<|\kappa|>}$ in the case of redshift bin $3.9 \leq z \leq 4.1$ compared to study of \citet{Becker2011}. It is worth noting that it is problematic to compare these values due to the curvature values depends on the data (noise, resolution) as well as on the method of noise treatment. In a preliminary analysis, we found that the main source of this discrepancy is the noisier dataset we used compared to the study of \citet{Becker2007}.
\subsection{Characteristic overdensities}
\begin{figure}
    \centering
    \includegraphics[width=0.45\columnwidth]{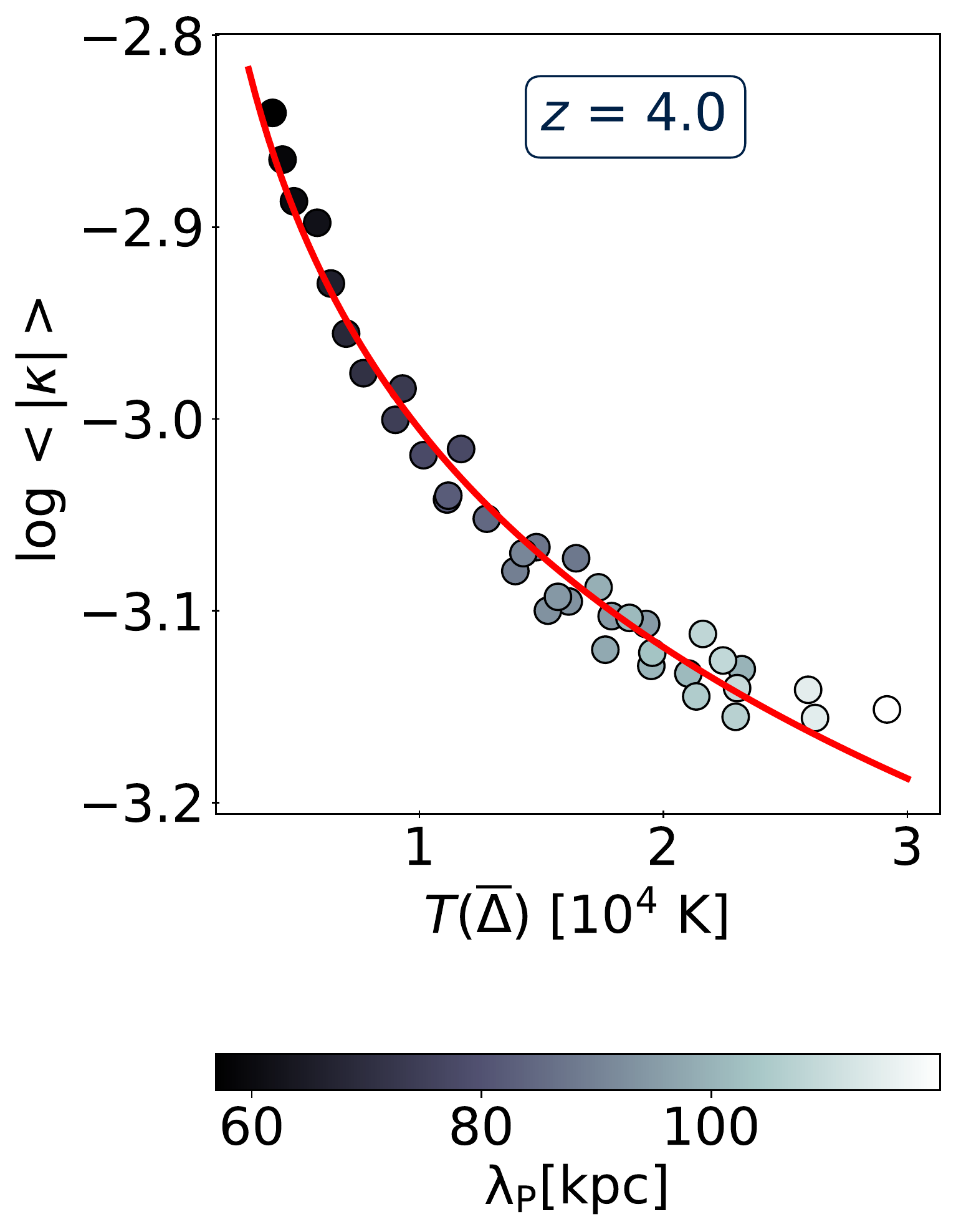}
    \includegraphics[width=0.45\columnwidth]{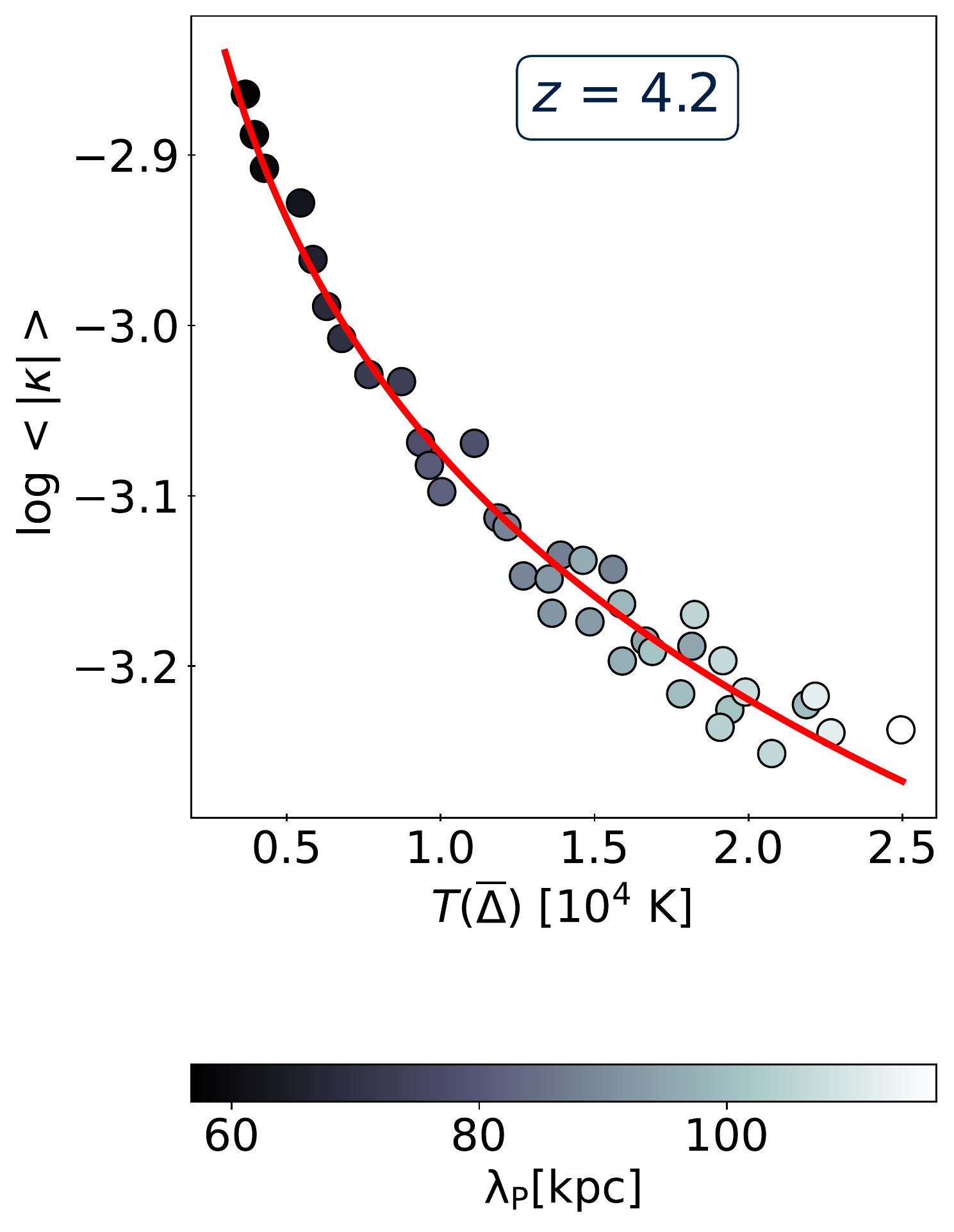}
    \caption{$\log{<|\kappa|>}$ as a function of $T(\overline{\Delta})$ for our simulations.}
    \label{fig:kappa_T}
\end{figure}

As was shown in \citep{Becker2011,Boera2014}, the $\log{<|\kappa|>}$ follows the tight relation with the gas temperature at the characteristic overdensity \citep{Padmanabhan2015}. The method to inferring the characteristic overdensities used in this study can be explained as follows:
\begin{enumerate}
\item We determined the $\log{<|\kappa|>}$ of the simulated spectra for each imput model. In this case, the final values of mean absolute curvature corresponds to the median of ${<|\kappa|>}$ from the 200 mock datasets generated for each input model.
\item For a given value of $\Delta$, we calculated the $T(\Delta)$ for each model using the $T_{0}$ and $\gamma$.
\item We plotted the values of $T(\Delta)$ versus $\log{<|\kappa|>}$ for each input model (Fig. \ref{fig:kappa_T}), and fit the relation using a power-law fit using the least squares method:
\begin{equation}
\log{<|\kappa|>}= - \left( \frac{T(\Delta)}{A} \right) ^{1/\alpha},
\label{char_overdensity}
\end{equation}
where $A$ and $\alpha$ are the free parameters. 
\item Subsequently, we varied the value of $\Delta$ in Eq. (\ref{char_overdensity}) and determined its value (by varying $A$ and $\alpha$), which corresponds to the best-fit.
\end{enumerate}
Note that the value of $\Delta$ obtained by the aforementioned approach is denote by $\overline{\Delta}$ and is defined as the 'characteristic overdensity' associated with the mean curvature \citep{Padmanabhan2015}. 

To quantify the amount of scatter in Fig. \ref{fig:kappa_T}, we determined the values of the characteristic overdensities and corresponding best-fitting parameters $A$ and $\alpha$ from the 5\,000 bootstrap realizations of the curvature values of the input models. These were used for the calculation of $T(\overline{\Delta})$ and also $T_{0}$ (see below). The final fits in Fig. \ref{fig:kappa_T} are based on the median values of the $\overline{\Delta}$ and corresponding best-fitting parameters $A$ and $\alpha$.

\subsection{Temperature at the characteristic overdensity}
\begin{figure}
    \centering
    \includegraphics[width=0.75\columnwidth]{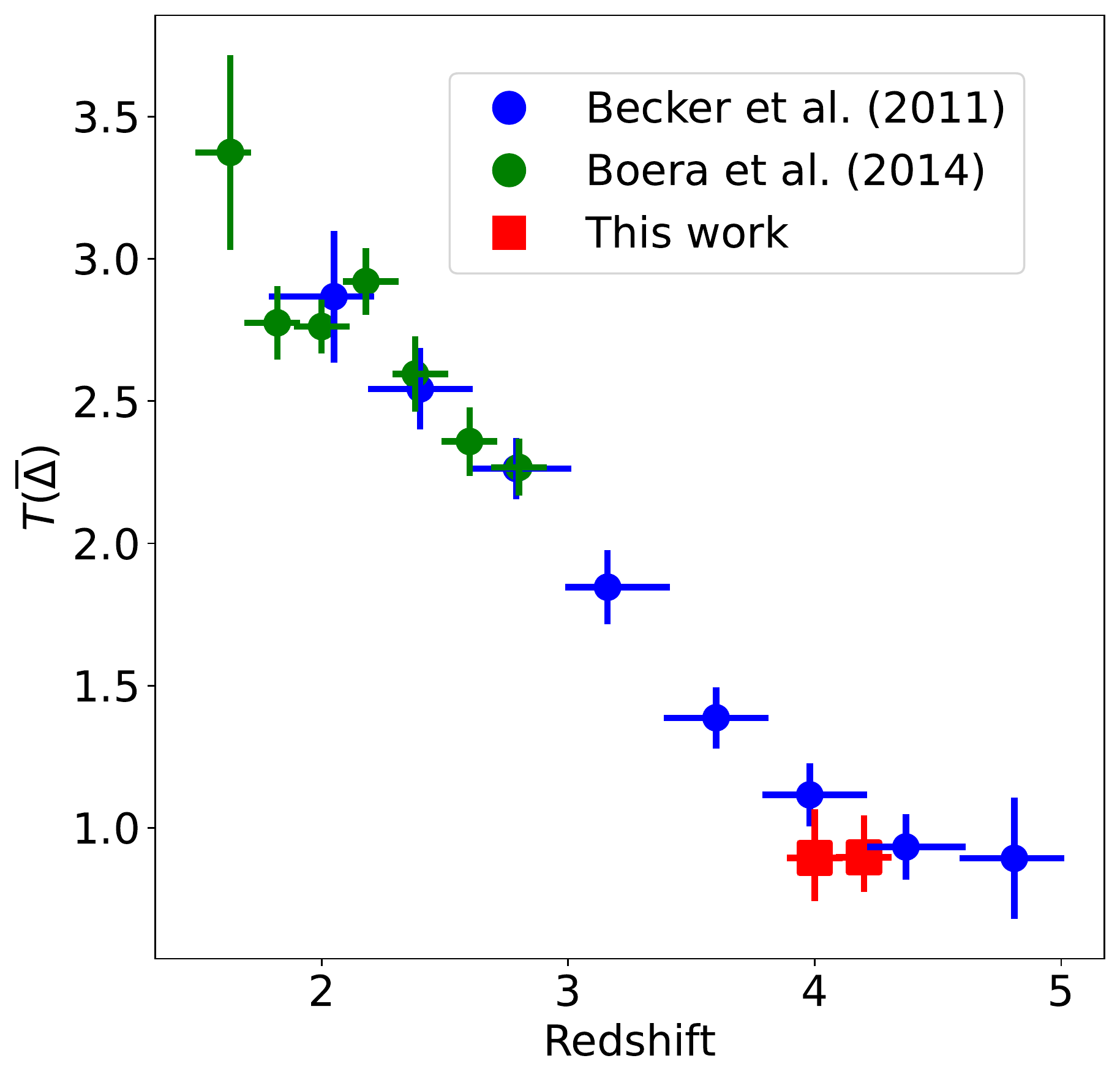}
    \caption{Comparison of the temperatures of the intergalactic medium at the optimal overdensity as a function of redshift obtained in this study and previously published ones.}
    \label{fig:T_delta}
\end{figure}

In the previous part of the study, we determined the free parameters $A$ and $\alpha$, which allow us to calculate the $T(\overline{\Delta})$ from the $\log{<|\kappa|>}$ of the QSO spectra for both redshift bins using Eq. (\ref{char_overdensity}). It is worth noting that in this case, we combine the values of $\overline{\Delta}$, $A$ and $\alpha$ with the $\log < |\kappa| >$ of the QSO spectra obtained by bootstrap method. Subsequently, the medians were used as the best estimates of $T(\overline{\Delta})$ and $T_{0}$ (see below). This approach also includes uncertainties which arose during the individual steps implemented in the analysis. The results show that our measurements are in a good agreement with the previous study by \citet{Becker2011}, and are depicted in Fig. \ref{fig:T_delta}.

\subsection{Temperature at the mean density}
We can convert the values of $T(\overline{\Delta})$ into $T_{0}$ using Eq. (\ref{asymptot_td_relation2}), which requires knowing the value of $\gamma$. Under the assumption of $\gamma = 1.4$, motivated by the evolution of this parameter predicted by the various UVB models, we determined the values of temperatures at mean density $T_{0} = 7893^{+1417}_{-1226}$\,K and $T_{0} = 8153^{+1224}_{-993}$\,K for redshift range of $3.9 \leq z \leq 4.1$ and $4.1 \leq z \leq 4.3$, respectively. All derived value of parameters are listed in Table \ref{tab:Results}.

\begin{table*}
	\begin{center}
	\caption{Values of the parameters determined for the investigated redshift ranges (column 1): total numbers of used 20\,Mpc/\boldmath{$h$} sections (column 2), associated characteristic overdensities (column 3), values of free parameters in Eq. (\ref{char_overdensity}) (columns 4 and 5), mean absolute curvature values (column 6), temperature measurements at the characteristic overdensity (column 7) and at mean density under the assumption of \boldmath{$\gamma = 1.4$} (column 8).}
	\label{tab:Results}
	\begin{tabular}{l c c c c c c c} \toprule
$z$ & $N$ & $\overline{\Delta}$ & $A$ & $\alpha$ & $<|\kappa|> \times 10^{-4}$ & $T(\overline{\Delta})$ & $T_{0}^{\gamma = 1.4}$ \\
\midrule
3.9 - 4.1 & 32 & 1.36 & $1.19\times10^{-5}$ & 18.67 & $10.28^{+0.63}_{-0.59}$ & $8943^{+1604}_{-1401}$ & $7893^{+1417}_{-1226}$ \\
4.1 - 4.3 & 15 & 1.27 & $4.24\times10^{-4}$ & 15.11 & $8.84^{+0.53}_{-0.54}$ & $8965^{+1370}_{-1111}$ & $8153^{+1224}_{-993}$ \\
\bottomrule
	\end{tabular}\\
	\end{center}
\end{table*}

\subsection{Comparison with previous studies}
The comparison of the results obtained in this study with previously published ones is shown in Fig. \ref{fig:models_Gaikwad}. The derived value of $T_{0}$ (within uncertainty) is consistent with that published by \citet{Walther2019} in case of both studied redshift bins. Comparing with the study of \citet{Becker2011}, when we rescaled their values assuming $\gamma = 1.4$, we obtained the similar value of $T_{0}$.

In case of the bin which corresponds to the higher redshift ($z = 4.2$) our results correspond to the results presented by \citet{Garzilli2017} and \citet{Boera_2019}. Note that due to similarity of our results and ones of the aforementioned study, the points in the Fig. \ref{fig:models_Gaikwad}, which corresponds to the \citet{Boera_2019} are overlapped with our values.

\subsection{Comparison with models}
\label{App_gaikwad}
We also compared the obtained results with predictions of five widely used UVB models with rescaled H\,{\sc i}, He\,{\sc i} and He\,{\sc ii} photoheating rates as presented in the study of \citet{gaikwad2021}. Based on the measurement of the thermal state of the IGM in the redshift range of $2 \leq z \leq 4$ determined using various statistics available in the literature (i.e. flux power spectrum, wavelet statistics, curvature statistics, and $b$-parameter probability distribution function) the authors found a good match between the shape of the observed $T_{0}$ and $\gamma$ evolution and that predicted by the UVB models with scaled photo-heating rates.

\begin{figure*}
\centering
	\includegraphics[width=0.75\columnwidth]{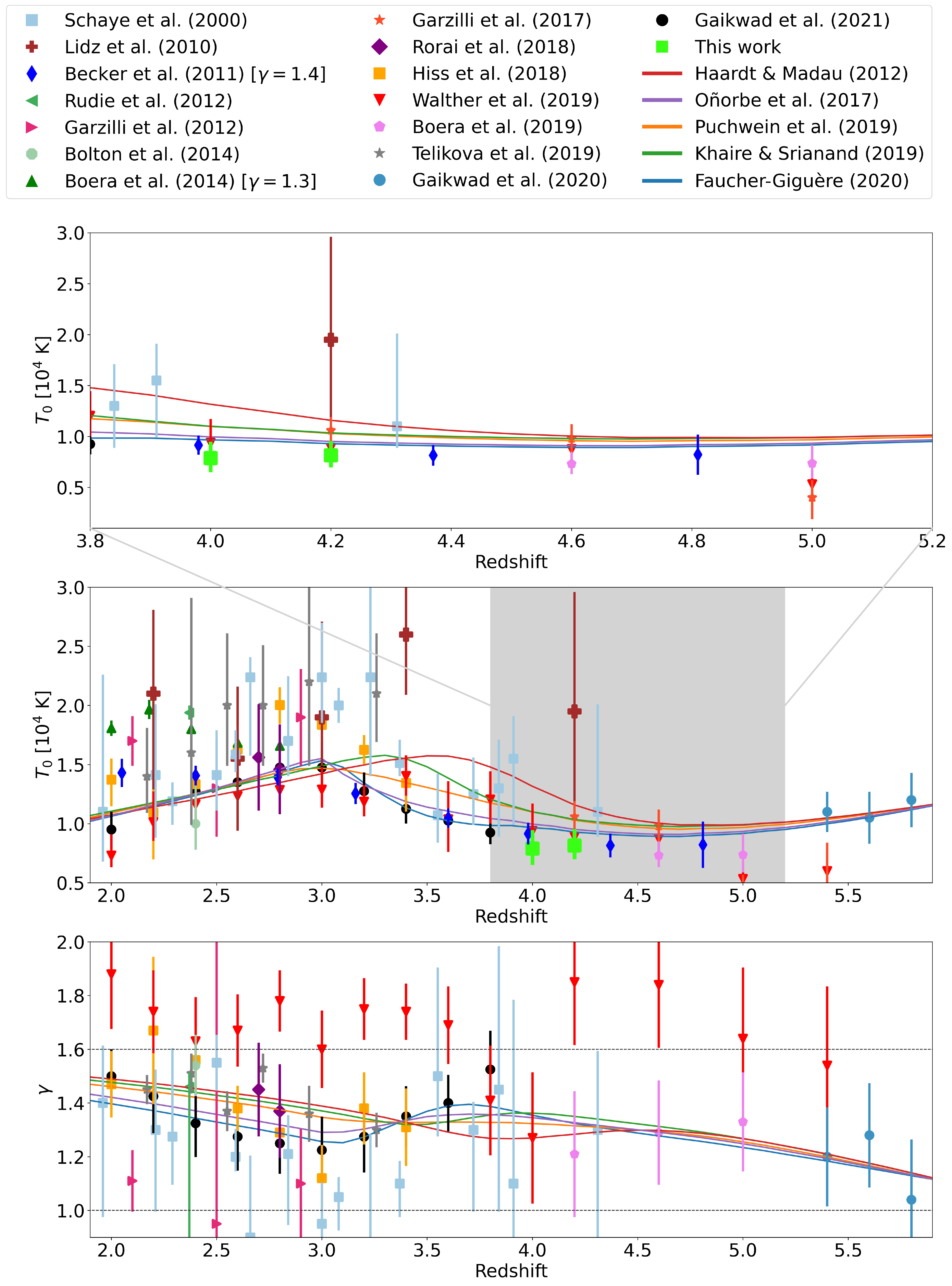}
    \caption{Comparison of the results obtained in this study with previously published ones and various models. We plotted the $T_{0}$ and $\gamma$ evolution for the UVB models of \citet{Haardt_2012}, \citet{Onorbe2017}, \citet{Khaire}, \citet{Puchwein2019}, and \citet{Faucher2020} with red, purple, green, orange, and blue line, respectively. The photoheating rates of the \citet{Onorbe2017}, \citet{Puchwein2019}, and \citet{Faucher2020} UVB models are scaled by a factors of 0.8, 0.9, and 0.7, respectively.}
    \label{fig:models_Gaikwad}
\end{figure*}

The rescaled models, as were presented in the aforementioned study, together with our results are showed in Fig. \ref{fig:models_Gaikwad}. In the case of the lower redshift bin, the determined value of $T_{0}$ is lower that predicted by the UVB models. On the other hand, in the case of the higher redshift bin, the  determined value of $T_{0}$ corresponds (within the error) with that predicted by the models of \citet{Onorbe2017} and \citet{Faucher2020}. It can be concluded that these results are consistent with the relatively late He\,{\sc ii} reionization in the aforementioned models.

\nocite{Rudie2012,Garzilli2012, Bolton2014,Gaikwad2020}
\section{CONCLUSIONS}
In this study, we applied the curvature method on a sample of 10~QSO spectra obtained by the Ultraviolet and Visual Echelle Spectrograph on the VLT/ESO to obtained the value of IGM temperature at a mean density. The main results could be summarized as follows:
\begin{itemize}
    \item Adopting the assumption of $\gamma = 1.4$, we determined the values of IGM temperatures at mean density $T_{0} = 7893^{+1417}_{-1226}$\,K and $T_{0} = 8153^{+1224}_{-993}$\,K for redshift range of $3.9 \leq z \leq 4.1$ and $4.1 \leq z \leq 4.3$, respectively.
    \item The value of $T_0$ that we derived from our $T(\overline{\Delta})$ starts to be largely independent of $\gamma$, with increasing $z$, because we have measured the temperature close to the mean density.
    \item Although the results show no strong temperature evolution over the studied redshift range, our measurements are consistent with the relatively late He\,{\sc ii} reionization presented in the \citet{Onorbe2017} and \citet{Faucher2020} models.
\end{itemize}

\begin{acknowledgement}
This research is based on the data products created from observations collected at the European Organisation for Astronomical Research in the Southern Hemisphere. The authors would also like to thank Jose O\~{n}orbe, Prakash Gaikwad and George Becker for the fruitful discussion, and to the anonymous referee for carefully reading our manuscript and for insightful comments and suggestions that improved the quality of this work.
\end{acknowledgement}


\bibliography{example}

\appendix

\section{Steps of the curvature calculation}
\label{Sec:app1}
For the purpose of reproducibility, in this section we describe the basic steps of the curvature calculation, which include the first and second derivatives and the Gaussian filter.

\begin{figure}
    \centering
    \includegraphics[width=0.4\columnwidth]{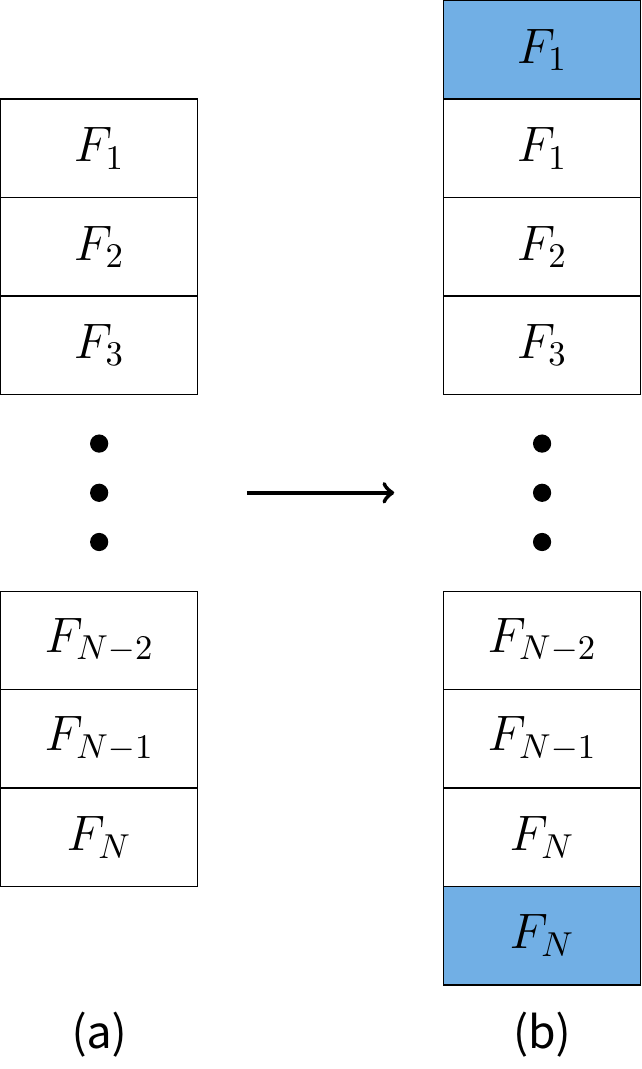}
    \caption{Input and output arrays for the derivative calculation.}
    \label{fig:der_fig}
\end{figure}

\subsection{First and second derivative}
Firstly, we create the new array in which the first and second values correspond to the first value of re-normalized flux. Similarly, the last two values correspond to the last value of the re-normalized flux array (see Fig. \ref{fig:der_fig}). 

Then, for the derivative calculation we used the centered difference approximations: 
\begin{equation}
    F'(x) = \frac{F(x+h) - F(x-h)}{2h}
\end{equation}
\begin{equation}
    F''(x) = \frac{F(x-h) - 2F(x) + F(x+h)}{h^2}
\end{equation}

\subsection{Gaussian filter}
As was mentioned before, the curvature method can be affected by the finite S/N of the
spectra. To solve this difficulty, we smoothed the flux using the Gaussian filter according to the following algorithm.
\begin{figure}
    \centering
    \includegraphics[width=\columnwidth]{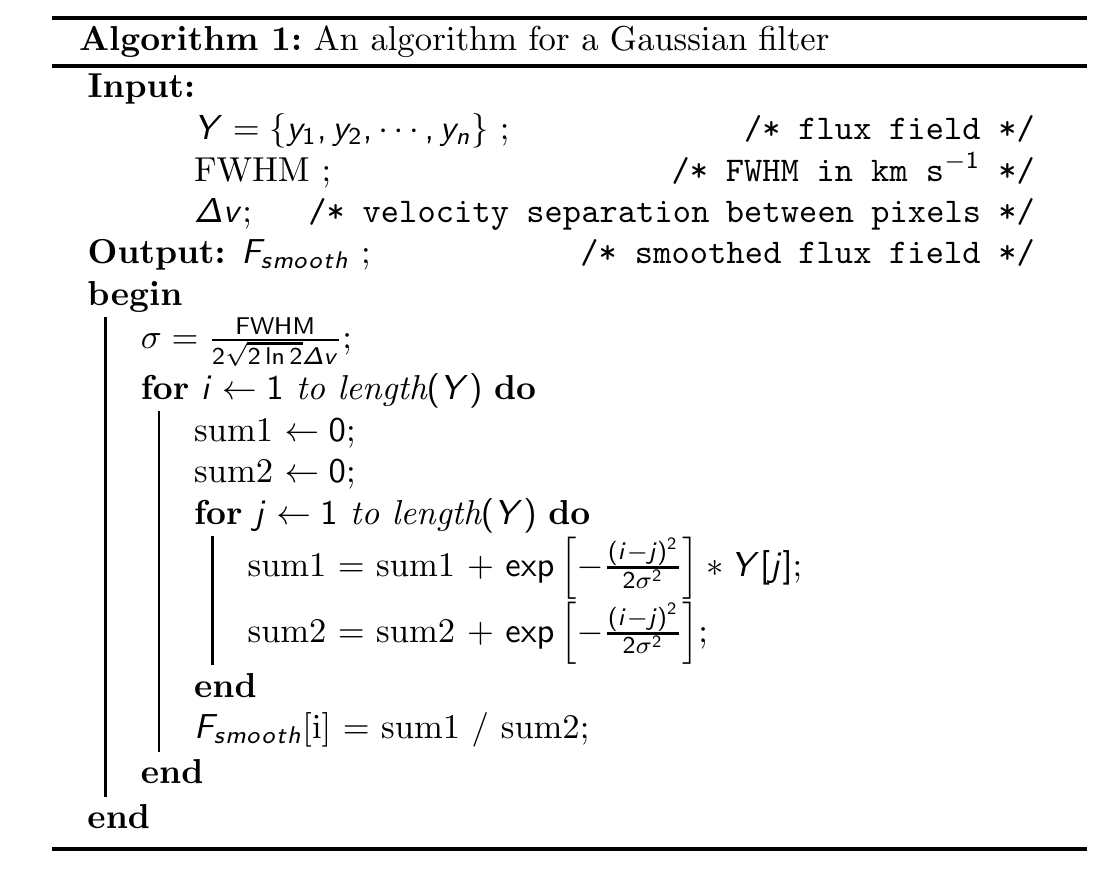}
    \label{fig:Algorithm}
\end{figure}

\end{document}